\documentstyle[12pt,moriond]{article}
\def\etal{{\it et~al\/}}
\def\Halpha{\hbox{$H\alpha$}}
\def\micron{\hbox{$\mu$m}}

\def\PsfigVersion{1.10}
\def\setDriver{\DvipsDriver} 
\ifx\undefined\psfig\else \fi
\let\LaTeXAtSign=\@
\let\@=\relax
\edef\psfigRestoreAt{\catcode`\@=\number\catcode`@\relax}
\catcode`\@=11\relax
\newwrite\@unused
\def\ps@typeout#1{{\let\protect\string\immediate\write\@unused{#1}}}

\def\DvipsDriver{
	\ps@typeout{psfig/tex \PsfigVersion -dvips}
\def\PsfigSpecials{\DvipsSpecials} 	\def\ps@dir{/}
\def\ps@predir{} }
\def\OzTeXDriver{
	\ps@typeout{psfig/tex \PsfigVersion -oztex}
	\def\PsfigSpecials{\OzTeXSpecials}
	\def\ps@dir{:}
	\def\ps@predir{:}
	\catcode`\^^J=5
}

\def\figurepath{./:}

\def\DoPaths#1{\expandafter\EachPath#1\stoplist}
\def\leer{}
\def\EachPath#1:#2\stoplist{
  \ExistsFile{#1}{\SearchedFile}
  \ifx#2\leer
  \else
    \expandafter\EachPath#2\stoplist
  \fi}
\def\ps@dir{/}
\def\ExistsFile#1#2{%
   \openin1=\ps@predir#1\ps@dir#2
   \ifeof1
       \closein1
   \else
       \closein1
        \ifx\ps@founddir\leer
           \edef\ps@founddir{#1}
        \fi
   \fi}
\def\get@dir#1{%
  \def\ps@founddir{}
  \def\SearchedFile{#1}
  \DoPaths\figurepath
}

\def\@nnil{\@nil}
\def\@empty{}
\def\@psdonoop#1\@@#2#3{}
\def\@psdo#1:=#2\do#3{\edef\@psdotmp{#2}\ifx\@psdotmp\@empty \else
    \expandafter\@psdoloop#2,\@nil,\@nil\@@#1{#3}\fi}
\def\@psdoloop#1,#2,#3\@@#4#5{\def#4{#1}\ifx #4\@nnil \else
       #5\def#4{#2}\ifx #4\@nnil \else#5\@ipsdoloop #3\@@#4{#5}\fi\fi}
\def\@ipsdoloop#1,#2\@@#3#4{\def#3{#1}\ifx #3\@nnil 
       \let\@nextwhile=\@psdonoop \else
      #4\relax\let\@nextwhile=\@ipsdoloop\fi\@nextwhile#2\@@#3{#4}}
\def\@tpsdo#1:=#2\do#3{\xdef\@psdotmp{#2}\ifx\@psdotmp\@empty \else
    \@tpsdoloop#2\@nil\@nil\@@#1{#3}\fi}
\def\@tpsdoloop#1#2\@@#3#4{\def#3{#1}\ifx #3\@nnil 
       \let\@nextwhile=\@psdonoop \else
      #4\relax\let\@nextwhile=\@tpsdoloop\fi\@nextwhile#2\@@#3{#4}}
\ifx\undefined\fbox
\newdimen\fboxrule
\newdimen\fboxsep
\newdimen\ps@tempdima
\newbox\ps@tempboxa
\long\def\fbox#1{\leavevmode\setbox\ps@tempboxa\hbox{#1}\ps@tempdima\fboxrule
    \advance\ps@tempdima \fboxsep \advance\ps@tempdima \dp\ps@tempboxa
   \hbox{\lower \ps@tempdima\hbox
  {\vbox{\hrule height \fboxrule
          \hbox{\vrule width \fboxrule \hskip\fboxsep
          \vbox{\vskip\fboxsep \box\ps@tempboxa\vskip\fboxsep}\hskip 
                 \fboxsep\vrule width \fboxrule}
                 \hrule height \fboxrule}}}}
\fi
\newread\ps@stream
\newif\ifnot@eof       
\newif\if@noisy        
\newif\if@atend        
\newif\if@psfile       
{\catcode`\%=12\global\gdef\epsf@start{
\def\epsf@PS{PS}
\def\epsf@getbb#1{%
\openin\ps@stream=\ps@predir#1
\ifeof\ps@stream\ps@typeout{Error, File #1 not found}\else
   {\not@eoftrue \chardef\other=12
    \def\do##1{\catcode`##1=\other}\dospecials \catcode`\ =10
    \loop
       \if@psfile
	  \read\ps@stream to \epsf@fileline
       \else{
	  \obeyspaces
          \read\ps@stream to \epsf@tmp\global\let\epsf@fileline\epsf@tmp}
       \fi
       \ifeof\ps@stream\not@eoffalse\else
       \if@psfile\else
       \expandafter\epsf@test\epsf@fileline:. \\%
       \fi
          \expandafter\epsf@aux\epsf@fileline:. \\%
       \fi
   \ifnot@eof\repeat
   }\closein\ps@stream\fi}%
\long\def\epsf@test#1#2#3:#4\\{\def\epsf@testit{#1#2}
			\ifx\epsf@testit\epsf@start\else
\ps@typeout{Warning! File does not start with `\epsf@start'.  It may not be a PostScript file.}
			\fi
			\@psfiletrue} 
{\catcode`\%=12\global\let\epsf@percent=
\long\def\epsf@aux#1#2:#3\\{\ifx#1\epsf@percent
   \def\epsf@testit{#2}\ifx\epsf@testit\epsf@bblit
	\@atendfalse
        \epsf@atend #3 . \\%
	\if@atend	
	   \if@verbose{
		\ps@typeout{psfig: found `(atend)'; continuing search}
	   }\fi
        \else
        \epsf@grab #3 . . . \\%
        \not@eoffalse
        \global\no@bbfalse
        \fi
   \fi\fi}%
\def\epsf@grab #1 #2 #3 #4 #5\\{%
   \global\def\epsf@llx{#1}\ifx\epsf@llx\empty
      \epsf@grab #2 #3 #4 #5 .\\\else
   \global\def\epsf@lly{#2}%
   \global\def\epsf@urx{#3}\global\def\epsf@ury{#4}\fi}%
\def\epsf@atendlit{(atend)} 
\def\epsf@atend #1 #2 #3\\{%
   \def\epsf@tmp{#1}\ifx\epsf@tmp\empty
      \epsf@atend #2 #3 .\\\else
   \ifx\epsf@tmp\epsf@atendlit\@atendtrue\fi\fi}

\chardef\psletter = 11 
\chardef\other = 12

\newif \ifdebug 
\newif\ifc@mpute 
\c@mputetrue 

\let\then = \relax
\def\r@dian{pt }
\let\r@dians = \r@dian
\let\dimensionless@nit = \r@dian
\let\dimensionless@nits = \dimensionless@nit
\def\internal@nit{sp }
\let\internal@nits = \internal@nit
\newif\ifstillc@nverging
\def \Mess@ge #1{\ifdebug \then \message {#1} \fi}

{ 
	\catcode `\@ = \psletter
	\gdef \nodimen {\expandafter \n@dimen \the \dimen}
	\gdef \term #1 #2 #3%
	       {\edef \t@ {\the #1}
		\edef \t@@ {\expandafter \n@dimen \the #2\r@dian}%
		\t@rm {\t@} {\t@@} {#3}%
	       }
	\gdef \t@rm #1 #2 #3%
	       {{%
		\count 0 = 0
		\dimen 0 = 1 \dimensionless@nit
		\dimen 2 = #2\relax
		\Mess@ge {Calculating term #1 of \nodimen 2}%
		\loop
		\ifnum	\count 0 < #1
		\then	\advance \count 0 by 1
			\Mess@ge {Iteration \the \count 0 \space}%
			\Multiply \dimen 0 by {\dimen 2}%
			\Mess@ge {After multiplication, term = \nodimen 0}%
			\Divide \dimen 0 by {\count 0}%
			\Mess@ge {After division, term = \nodimen 0}%
		\repeat
		\Mess@ge {Final value for term #1 of 
				\nodimen 2 \space is \nodimen 0}%
		\xdef \Term {#3 = \nodimen 0 \r@dians}%
		\aftergroup \Term
	       }}
	\catcode `\p = \other
	\catcode `\t = \other
	\gdef \n@dimen #1pt{#1} 
}

\def \Divide #1by #2{\divide #1 by #2} 

\def \Multiply #1by #2
       {{
	\count 0 = #1\relax
	\count 2 = #2\relax
	\count 4 = 65536
	\Mess@ge {Before scaling, count 0 = \the \count 0 \space and
			count 2 = \the \count 2}%
	\ifnum	\count 0 > 32767 
	\then	\divide \count 0 by 4
		\divide \count 4 by 4
	\else	\ifnum	\count 0 < -32767
		\then	\divide \count 0 by 4
			\divide \count 4 by 4
		\else
		\fi
	\fi
	\ifnum	\count 2 > 32767 
	\then	\divide \count 2 by 4
		\divide \count 4 by 4
	\else	\ifnum	\count 2 < -32767
		\then	\divide \count 2 by 4
			\divide \count 4 by 4
		\else
		\fi
	\fi
	\multiply \count 0 by \count 2
	\divide \count 0 by \count 4
	\xdef \product {#1 = \the \count 0 \internal@nits}%
	\aftergroup \product
       }}

\def\r@duce{\ifdim\dimen0 > 90\r@dian \then   
		\multiply\dimen0 by -1
		\advance\dimen0 by 180\r@dian
		\r@duce
	    \else \ifdim\dimen0 < -90\r@dian \then  
		\advance\dimen0 by 360\r@dian
		\r@duce
		\fi
	    \fi}

\def\Sine#1%
       {{%
	\dimen 0 = #1 \r@dian
	\r@duce
	\ifdim\dimen0 = -90\r@dian \then
	   \dimen4 = -1\r@dian
	   \c@mputefalse
	\fi
	\ifdim\dimen0 = 90\r@dian \then
	   \dimen4 = 1\r@dian
	   \c@mputefalse
	\fi
	\ifdim\dimen0 = 0\r@dian \then
	   \dimen4 = 0\r@dian
	   \c@mputefalse
	\fi
	\ifc@mpute \then
		\divide\dimen0 by 180
		\dimen0=3.141592654\dimen0
		\dimen 2 = 3.1415926535897963\r@dian 
		\divide\dimen 2 by 2 
		\Mess@ge {Sin: calculating Sin of \nodimen 0}%
		\count 0 = 1 
		\dimen 2 = 1 \r@dian 
		\dimen 4 = 0 \r@dian 
		\loop
			\ifnum	\dimen 2 = 0 
			\then	\stillc@nvergingfalse 
			\else	\stillc@nvergingtrue
			\fi
			\ifstillc@nverging 
			\then	\term {\count 0} {\dimen 0} {\dimen 2}%
				\advance \count 0 by 2
				\count 2 = \count 0
				\divide \count 2 by 2
				\ifodd	\count 2 
				\then	\advance \dimen 4 by \dimen 2
				\else	\advance \dimen 4 by -\dimen 2
				\fi
		\repeat
	\fi		
			\xdef \sine {\nodimen 4}%
       }}

\def\Cosine#1{\ifx\sine\UnDefined\edef\Savesine{\relax}\else
		             \edef\Savesine{\sine}\fi
	{\dimen0=#1\r@dian\advance\dimen0 by 90\r@dian
	 \Sine{\nodimen 0}
	 \xdef\cosine{\sine}
	 \xdef\sine{\Savesine}}}	      

\def\psdraft{
	\def\@psdraft{0}
}
\def\psfull{
	\def\@psdraft{100}
}

\psfull

\newif\if@scalefirst
\def\psscalefirst{\@scalefirsttrue}
\def\psrotatefirst{\@scalefirstfalse}
\psrotatefirst

\newif\if@draftbox
\def\psnodraftbox{
	\@draftboxfalse
}
\def\psdraftbox{
	\@draftboxtrue
}
\@draftboxtrue

\newif\if@prologfile
\newif\if@postlogfile
\def\pssilent{
	\@noisyfalse
}
\def\psnoisy{
	\@noisytrue
}
\psnoisy
\newif\if@bbllx
\newif\if@bblly
\newif\if@bburx
\newif\if@bbury
\newif\if@height
\newif\if@width
\newif\if@rheight
\newif\if@rwidth
\newif\if@angle
\newif\if@clip
\newif\if@verbose
\def\@p@@sclip#1{\@cliptrue}
\newif\if@decmpr
\def\@p@@sfigure#1{\def\@p@sfile{null}\def\@p@sbbfile{null}\@decmprfalse
   \openin1=\ps@predir#1
   \ifeof1
	\closein1
	\get@dir{#1}
	\ifx\ps@founddir\leer
		\openin1=\ps@predir#1.bb
		\ifeof1
			\closein1
			\get@dir{#1.bb}
			\ifx\ps@founddir\leer
				\ps@typeout{Can't find #1 in \figurepath}
			\else
				\@decmprtrue
				\def\@p@sfile{\ps@founddir\ps@dir#1}
				\def\@p@sbbfile{\ps@founddir\ps@dir#1.bb}
			\fi
		\else
			\closein1
			\@decmprtrue
			\def\@p@sfile{#1}
			\def\@p@sbbfile{#1.bb}
		\fi
	\else
		\def\@p@sfile{\ps@founddir\ps@dir#1}
		\def\@p@sbbfile{\ps@founddir\ps@dir#1}
	\fi
   \else
	\closein1
	\def\@p@sfile{#1}
	\def\@p@sbbfile{#1}
   \fi
}
\def\@p@@sfile#1{\@p@@sfigure{#1}}
\def\@p@@sbbllx#1{
		\@bbllxtrue
		\dimen100=#1
		\edef\@p@sbbllx{\number\dimen100}
}
\def\@p@@sbblly#1{
		\@bbllytrue
		\dimen100=#1
		\edef\@p@sbblly{\number\dimen100}
}
\def\@p@@sbburx#1{
		\@bburxtrue
		\dimen100=#1
		\edef\@p@sbburx{\number\dimen100}
}
\def\@p@@sbbury#1{
		\@bburytrue
		\dimen100=#1
		\edef\@p@sbbury{\number\dimen100}
}
\def\@p@@sheight#1{
		\@heighttrue
		\dimen100=#1
   		\edef\@p@sheight{\number\dimen100}
}
\def\@p@@swidth#1{
		\@widthtrue
		\dimen100=#1
		\edef\@p@swidth{\number\dimen100}
}
\def\@p@@srheight#1{
		\@rheighttrue
		\dimen100=#1
		\edef\@p@srheight{\number\dimen100}
}
\def\@p@@srwidth#1{
		\@rwidthtrue
		\dimen100=#1
		\edef\@p@srwidth{\number\dimen100}
}
\def\@p@@sangle#1{
		\@angletrue
		\edef\@p@sangle{#1} 
}
\def\@p@@ssilent#1{ 
		\@verbosefalse
}
\def\@p@@sprolog#1{\@prologfiletrue\def\@prologfileval{#1}}
\def\@p@@spostlog#1{\@postlogfiletrue\def\@postlogfileval{#1}}
\def\@cs@name#1{\csname #1\endcsname}
\def\@setparms#1=#2,{\@cs@name{@p@@s#1}{#2}}
\def\ps@init@parms{
		\@bbllxfalse \@bbllyfalse
		\@bburxfalse \@bburyfalse
		\@heightfalse \@widthfalse
		\@rheightfalse \@rwidthfalse
		\def\@p@sbbllx{}\def\@p@sbblly{}
		\def\@p@sbburx{}\def\@p@sbbury{}
		\def\@p@sheight{}\def\@p@swidth{}
		\def\@p@srheight{}\def\@p@srwidth{}
		\def\@p@sangle{0}
		\def\@p@sfile{} \def\@p@sbbfile{}
		\def\@p@scost{10}
		\def\@sc{}
		\@prologfilefalse
		\@postlogfilefalse
		\@clipfalse
		\if@noisy
			\@verbosetrue
		\else
			\@verbosefalse
		\fi
}
\def\parse@ps@parms#1{
	 	\@psdo\@psfiga:=#1\do
		   {\expandafter\@setparms\@psfiga,}}
\newif\ifno@bb
\def\bb@missing{
	\if@verbose{
		\ps@typeout{psfig: searching \@p@sbbfile \space  for bounding box}
	}\fi
	\no@bbtrue
	\epsf@getbb{\@p@sbbfile}
        \ifno@bb \else \bb@cull\epsf@llx\epsf@lly\epsf@urx\epsf@ury\fi
}	
\def\bb@cull#1#2#3#4{
	\dimen100=#1 bp\edef\@p@sbbllx{\number\dimen100}
	\dimen100=#2 bp\edef\@p@sbblly{\number\dimen100}
	\dimen100=#3 bp\edef\@p@sbburx{\number\dimen100}
	\dimen100=#4 bp\edef\@p@sbbury{\number\dimen100}
	\no@bbfalse
}
\newdimen\p@intvaluex
\newdimen\p@intvaluey
\def\rotate@#1#2{{\dimen0=#1 sp\dimen1=#2 sp
		  \global\p@intvaluex=\cosine\dimen0
		  \dimen3=\sine\dimen1
		  \global\advance\p@intvaluex by -\dimen3
		  \global\p@intvaluey=\sine\dimen0
		  \dimen3=\cosine\dimen1
		  \global\advance\p@intvaluey by \dimen3
		  }}
\def\compute@bb{
		\no@bbfalse
		\if@bbllx \else \no@bbtrue \fi
		\if@bblly \else \no@bbtrue \fi
		\if@bburx \else \no@bbtrue \fi
		\if@bbury \else \no@bbtrue \fi
		\ifno@bb \bb@missing \fi
		\ifno@bb \ps@typeout{FATAL ERROR: no bb supplied or found}
			\no-bb-error
		\fi
		\count203=\@p@sbburx
		\count204=\@p@sbbury
		\advance\count203 by -\@p@sbbllx
		\advance\count204 by -\@p@sbblly
		\edef\ps@bbw{\number\count203}
		\edef\ps@bbh{\number\count204}
		\if@angle 
			\Sine{\@p@sangle}\Cosine{\@p@sangle}
	        	{\dimen100=\maxdimen\xdef\r@p@sbbllx{\number\dimen100}
					    \xdef\r@p@sbblly{\number\dimen100}
			                    \xdef\r@p@sbburx{-\number\dimen100}
					    \xdef\r@p@sbbury{-\number\dimen100}}
                        \def\minmaxtest{
			   \ifnum\number\p@intvaluex<\r@p@sbbllx
			      \xdef\r@p@sbbllx{\number\p@intvaluex}\fi
			   \ifnum\number\p@intvaluex>\r@p@sbburx
			      \xdef\r@p@sbburx{\number\p@intvaluex}\fi
			   \ifnum\number\p@intvaluey<\r@p@sbblly
			      \xdef\r@p@sbblly{\number\p@intvaluey}\fi
			   \ifnum\number\p@intvaluey>\r@p@sbbury
			      \xdef\r@p@sbbury{\number\p@intvaluey}\fi
			   }
			\rotate@{\@p@sbbllx}{\@p@sbblly}
			\minmaxtest
			\rotate@{\@p@sbbllx}{\@p@sbbury}
			\minmaxtest
			\rotate@{\@p@sbburx}{\@p@sbblly}
			\minmaxtest
			\rotate@{\@p@sbburx}{\@p@sbbury}
			\minmaxtest
			\edef\@p@sbbllx{\r@p@sbbllx}\edef\@p@sbblly{\r@p@sbblly}
			\edef\@p@sbburx{\r@p@sbburx}\edef\@p@sbbury{\r@p@sbbury}
		\fi
		\count203=\@p@sbburx
		\count204=\@p@sbbury
		\advance\count203 by -\@p@sbbllx
		\advance\count204 by -\@p@sbblly
		\edef\@bbw{\number\count203}
		\edef\@bbh{\number\count204}
}
\def\in@hundreds#1#2#3{\count240=#2 \count241=#3
		     \count100=\count240	
		     \divide\count100 by \count241
		     \count101=\count100
		     \multiply\count101 by \count241
		     \advance\count240 by -\count101
		     \multiply\count240 by 10
		     \count101=\count240	
		     \divide\count101 by \count241
		     \count102=\count101
		     \multiply\count102 by \count241
		     \advance\count240 by -\count102
		     \multiply\count240 by 10
		     \count102=\count240	
		     \divide\count102 by \count241
		     \count200=#1\count205=0
		     \count201=\count200
			\multiply\count201 by \count100
		 	\advance\count205 by \count201
		     \count201=\count200
			\divide\count201 by 10
			\multiply\count201 by \count101
			\advance\count205 by \count201
		     \count201=\count200
			\divide\count201 by 100
			\multiply\count201 by \count102
			\advance\count205 by \count201
		     \edef\@result{\number\count205}
}
\def\compute@wfromh{
		\in@hundreds{\@p@sheight}{\@bbw}{\@bbh}
		\edef\@p@swidth{\@result}
}
\def\compute@hfromw{
	        \in@hundreds{\@p@swidth}{\@bbh}{\@bbw}
		\edef\@p@sheight{\@result}
}
\def\compute@handw{
		\if@height 
			\if@width
			\else
				\compute@wfromh
			\fi
		\else 
			\if@width
				\compute@hfromw
			\else
				\edef\@p@sheight{\@bbh}
				\edef\@p@swidth{\@bbw}
			\fi
		\fi
}
\def\compute@resv{
		\if@rheight \else \edef\@p@srheight{\@p@sheight} \fi
		\if@rwidth \else \edef\@p@srwidth{\@p@swidth} \fi
}
\def\compute@sizes{
	\compute@bb
	\if@scalefirst\if@angle
	\if@width
	   \in@hundreds{\@p@swidth}{\@bbw}{\ps@bbw}
	   \edef\@p@swidth{\@result}
	\fi
	\if@height
	   \in@hundreds{\@p@sheight}{\@bbh}{\ps@bbh}
	   \edef\@p@sheight{\@result}
	\fi
	\fi\fi
	\compute@handw
	\compute@resv}
\def\OzTeXSpecials{
	\special{empty.ps /@isp {true} def}
	\special{empty.ps \@p@swidth \space \@p@sheight \space
			\@p@sbbllx \space \@p@sbblly \space
			\@p@sbburx \space \@p@sbbury \space
			startTexFig \space }
	\if@clip{
		\if@verbose{
			\ps@typeout{(clip)}
		}\fi
		\special{empty.ps doclip \space }
	}\fi
	\if@angle{
		\if@verbose{
			\ps@typeout{(rotate)}
		}\fi
		\special {empty.ps \@p@sangle \space rotate \space} 
	}\fi
	\if@prologfile
	    \special{\@prologfileval \space } \fi
	\if@decmpr{
		\if@verbose{
			\ps@typeout{psfig: Compression not available
			in OzTeX version \space }
		}\fi
	}\else{
		\if@verbose{
			\ps@typeout{psfig: including \@p@sfile \space }
		}\fi
		\special{epsf=\ps@predir\@p@sfile \space }
	}\fi
	\if@postlogfile
	    \special{\@postlogfileval \space } \fi
	\special{empty.ps /@isp {false} def}
}
\def\DvipsSpecials{
	\special{ps::[begin] 	\@p@swidth \space \@p@sheight \space
			\@p@sbbllx \space \@p@sbblly \space
			\@p@sbburx \space \@p@sbbury \space
			startTexFig \space }
	\if@clip{
		\if@verbose{
			\ps@typeout{(clip)}
		}\fi
		\special{ps:: doclip \space }
	}\fi
	\if@angle
		\if@verbose{
			\ps@typeout{(clip)}
		}\fi
		\special {ps:: \@p@sangle \space rotate \space} 
	\fi
	\if@prologfile
	    \special{ps: plotfile \@prologfileval \space } \fi
	\if@decmpr{
		\if@verbose{
			\ps@typeout{psfig: including \@p@sfile.Z \space }
		}\fi
		\special{ps: plotfile "`zcat \@p@sfile.Z" \space }
	}\else{
		\if@verbose{
			\ps@typeout{psfig: including \@p@sfile \space }
		}\fi
		\special{ps: plotfile \@p@sfile \space }
	}\fi
	\if@postlogfile
	    \special{ps: plotfile \@postlogfileval \space } \fi
	\special{ps::[end] endTexFig \space }
}
\def\psfig#1{\vbox {
	\ps@init@parms
	\parse@ps@parms{#1}
	\compute@sizes
	\ifnum\@p@scost<\@psdraft{
		\PsfigSpecials 
		\vbox to \@p@srheight sp{
			\hbox to \@p@srwidth sp{
				\hss
			}
		\vss
		}
	}\else{
		\if@draftbox{		
			\hbox{\fbox{\vbox to \@p@srheight sp{
			\vss
			\hbox to \@p@srwidth sp{ \hss 
			 \hss }
			\vss
			}}}
		}\else{
			\vbox to \@p@srheight sp{
			\vss
			\hbox to \@p@srwidth sp{\hss}
			\vss
			}
		}\fi

	}\fi
}}
\psfigRestoreAt
\setDriver
\let\@=\LaTeXAtSign

\begin{document}
\heading{Deep Near-Infrared Surveys --- Understanding
Galaxy Evolution at \boldmath $z>1$}

\author{K. Glazebrook} {Anglo-Australian Observatory, P.O. Box 296,
       Epping, NSW 2121, AUSTRALIA.}  {\ }

\begin{moriondabstract}

Deep near-infrared (NIR) surveys are critical to our current, and even more
to our future, understanding of galaxy evolution in the early universe.
In this review I will be discussing the relevance of deep NIR surveys 
and looking at the information provided by
different types of survey: broad-band imaging, 
spectroscopic observations and narrow-band imaging.  
In particular I will be looking at the
future possibilities for faint galaxy work provided by forthcoming, innovative,
NIR instrumentation being developed for the next generation of 8\,m telescopes.

\end{moriondabstract}

\section{Why The Near-Infrared?}

The initial interest in the NIR for studies of galaxy
evolution came from the realisation that at low redshift ($z<1$) the NIR
K-corrections for different morphological classes of galaxy are very
similar (Figure~1). Thus by selecting in the NIR the morphological
mix of the sample, {\em i.e.} the ratio of galaxies with young
and old stellar populations, would be insensitive to redshift
if there was no evolution. Thus any evolutionary signature would
be more clearly seen.  

\begin{figure} 
\hskip 3cm\psfig{file=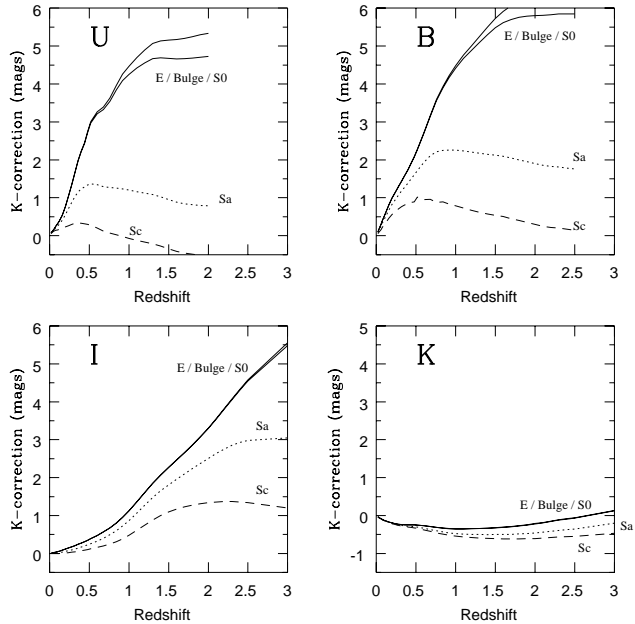,width=9cm,angle=-90}
\caption{ \protect\small K-corrections in ultraviolet through optical to NIR
broad-bands for galaxies with elliptical (old stellar populations) to Sc
(star-forming populations) spectral energy distributions (SEDs). It can be seen
that in the $K$-band the band-shifting effect is very similar for the different
SEDs. In contrast in the $U$-band elliptical SEDs are dimmed
by many magnitudes relative to later types even at moderate redshifts.
}
\end{figure}

This led to many early NIR surveys at relatively bright magnitudes (e.g.
\cite{GARD93},\cite{KGB94}) which confirmed a general picture of
very little evolution in early-type galaxies along with a general
increase in the space density of star-forming systems at low redshift.
Such surveys are reviewed elsewhere in these Proceedings \cite{GARD}.

Here we will consider deeper surveys, with IR magnitudes typically
$K>19$--20 in which a significant number of galaxies lie beyond a
redshift of unity. At these redshifts the prominent optical emission
and absorption spectra features are shifted into the NIR. For example
the [OII] and Ca H and K lines are in the $J$-band for $z>1.6$ and for
$z>2$ the \Halpha\ line lies in the $K$-band. This has so far
provided a natural limit for redshift surveys in the optical, the
deepest extending to $z=1.5$ \cite{COWIE95}. For $z=1$--4
the K-corrections
will no longer be as uniform as in the brighter surveys
(see Figure~1) but there is still less
variation than in optical and UV bands. Also the galaxies selected
in deep NIR surveys are being observed in rest-frame optical light,
thus their NIR properties can be compared with the well-studied 
optical properties of local galaxies. Thus evolutionary signatures
of rest-frame optical parameters (colours, luminosities, line
strengths) etc. can be determined in a fairly model-free manner.

The problem of course with studying very faint sources in the NIR
is increased brightness of the night sky. This is mainly due to the
emission of very many narrow OH lines (Figure~2). The key concept
behind the next generation of NIR instruments is to filter
out these lines and reduce this background.

\begin{figure} 
\hskip 2cm\psfig{file=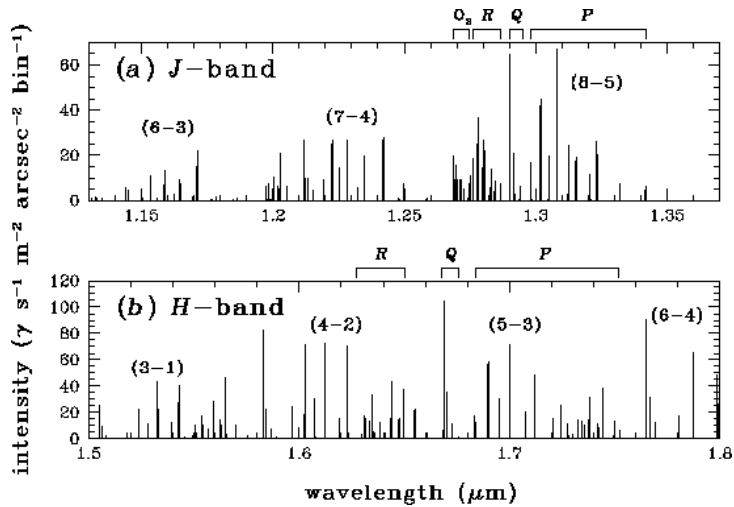,height=7cm}
\caption{\protect\small Night sky spectrum in $J$ and $H$-bands. 95\% of the
background comes from narrow OH emission lines.}
\end{figure}

\section{Imaging Surveys}

Most activity in deep NIR cosmological work has concentrated on the
deep $K$-band counts \cite{DJO}\cite{MAC}. 
Much has been made of their significance --- in
particular the turnover (or lack thereof) in the faint end and it's
implications for the cosmological geometry. 
This has been because the $K$-band counts
have been found to lie closer to the prediction of a Robertson-Walker
metric than in any other band (optical/Far IR/radio) and $K$-band
counts are `insensitive to evolution.' We must bear in mind though
that beyond $K\sim 19$ the galaxies are at high-redshift and being
observed in rest-frame optical and the evolutionary effects of stellar
populations are not necessarily negligible. Moreover these effects
tend to be of $O(z)$ due to their time dependence and so will always
win out over cosmological effects of order $O(z^2)$. Also any
evolution in the space density of galaxies (after all galaxies {\em must}
form at some redshift) will affect the faint end $K$-band counts. 
\begin{figure}[t]
\begin{tabular}{cc}
\psfig{file=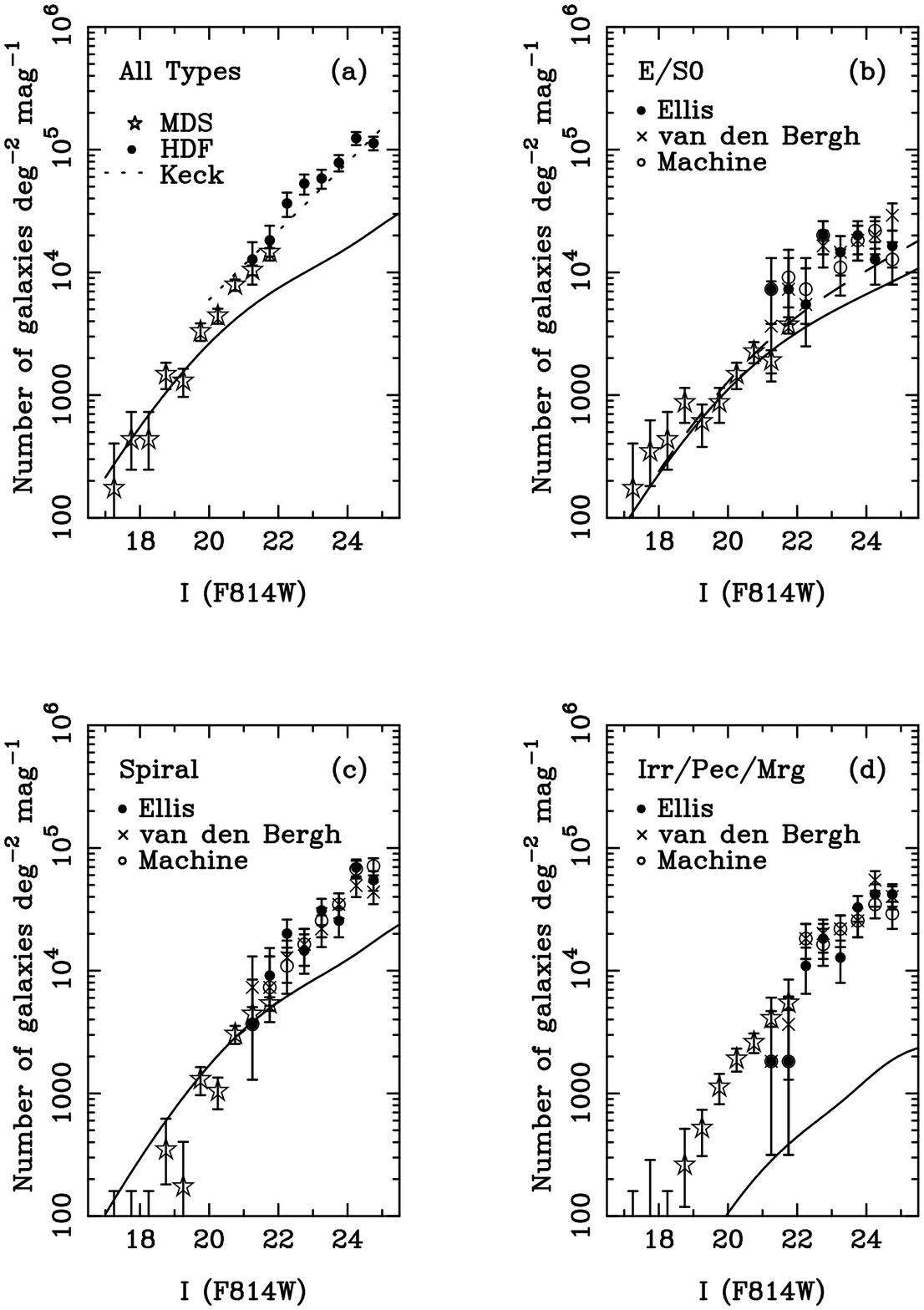,width=8.5cm,rwidth=8cm} &
\vbox{\psfig{file=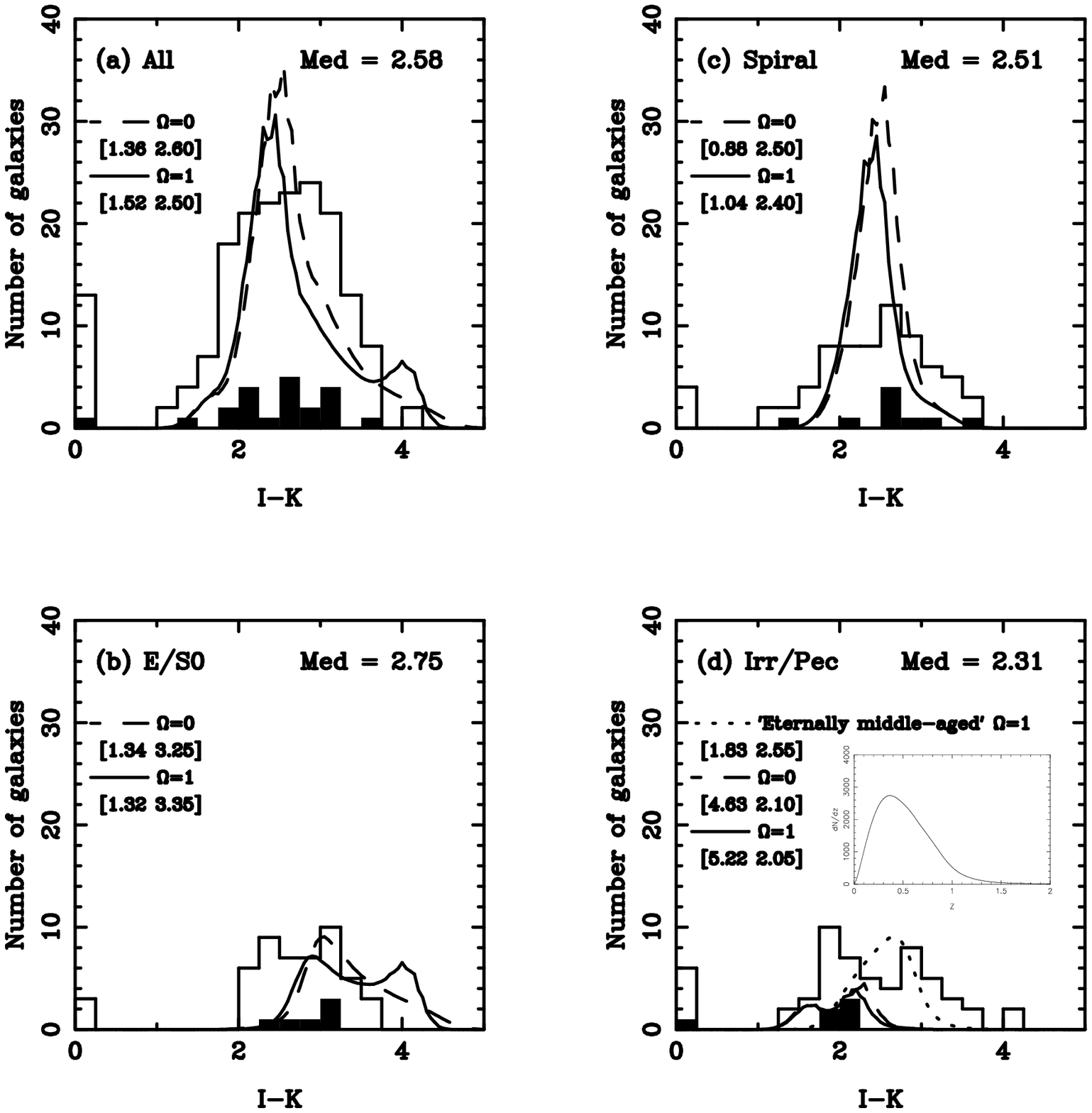,width=8cm,angle=-90}\vskip 1cm}\\
\end{tabular}
\caption{\protect\small
Left: The number-magnitude relations for morphologically segregated (by visual
and automated classifiers) samples of galaxies
from the MDS (stars) and HDF (circles). The
no-evolution $\Omega=1$ curves from \protect\cite{KGB95}, are superposed. The dashed line
on the E/S0 diagram shows the effect of assuming $\Omega=0.1$. \hfill\break 
Right: $I-K$ colour distributions from
\protect\cite{MDSPHYS} of morphologically  resolved classes at $I<22$ compared with
predictions of pure luminosity evolution models. The colours indicate a population of
blue compact objects contaminating the `ellipticals' category, and also show that the
`irregular/peculiar' category are not consistent with pure young starburst colours.
}
\end{figure}
Perhaps more useful than the broad-brush of total counts are the
morphologically resolved counts that have been done with the
{\it Hubble Space Telescope} \cite{HDF},\cite{DRIVER},\cite{KGB95}. 
These show a picture of evolution where ellipticals evolve slowly
(though there is an evolving population of blue compact galaxies \cite{SCHADE},
\cite{MDSPHYS}), spiral galaxies show a more rapid evolution (with significant
changes in morphological appearance \cite{VDB}) and there
is dramatic evolution in populations of morphologically irregular
and peculiar galaxies, including mergers (Figure 3). It should be
noted that differential evolution is found in the deep surveys {
\em independently}
of the known uncertainties in the local luminosity function (for
a discussion thereof see \cite{GARD}).

It is clear that the next step is to combine high-resolution with
multi-colour photometry to allow the study of the evolution of
stellar populations {\em within} galaxies. This will allow us
to answer questions such as the epochs of bulge and disk formation,
where the star-formation takes place in different parts of galaxies
as a function of epoch, etc. An early example is the
pioneering work of Abraham \etal\ (Figure~4) using optical colours
from the Hubble Deep Field (HDF).

\begin{figure}
\psfig{file=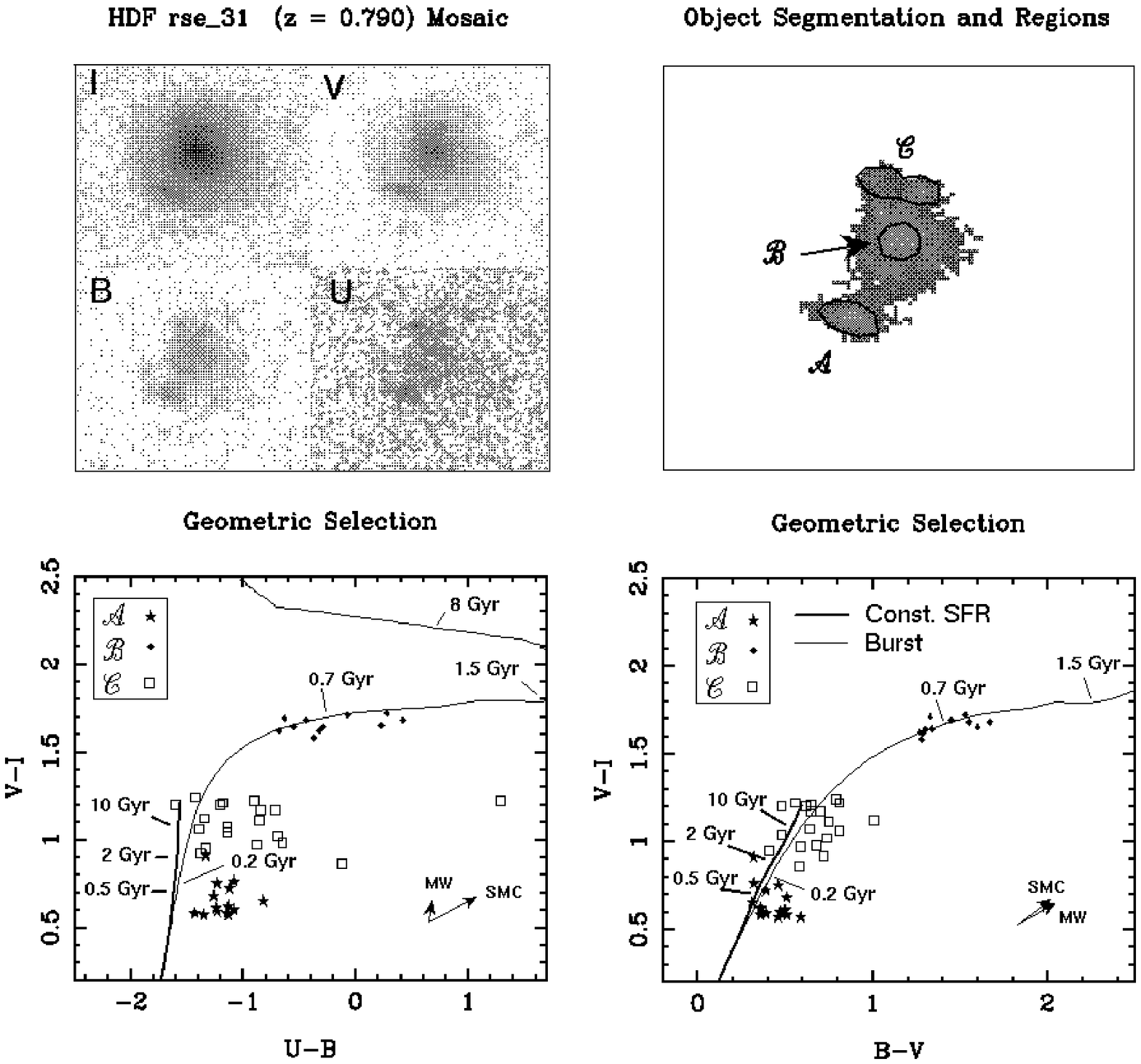,width=\hsize}
\caption{\protect\small Morphologically resolved spectrophotometry from Abraham
\etal\ (work in progress). This is just {\em one} galaxy (at $z=0.79$)
from a sample of
dozens in the Hubble Deep Field and morphologically appears `proto-spiral.'
Selected morphlogical regions of the galaxy
($\cal A$, $\cal B$ and $\cal C$) are segmented from the image and their pixels plotted
in $U-B$, $B-V$ and $V-I$. The thin and thick lines show
the tracks of single-epoch burst and continuous star-forming stellar
populations from \protect\cite{BC96} of 40\% solar metallicity and
with ages at various points indicated. The arrows, at the bottom right,
indicate SMC and Milky Way
reddening laws for $A_B=0.4$ mag. 
It can be seen that to first order the colours lie along
a sequence of progressively older stellar populations, a very encouraging
agreement with the model colours. By isolating the populations morphologically
the colours show less dispersion than do global colours of unresolved objects.
It can also be seen that the `proto-bulge' ($\cal B$) 
is the reddest component and the $B-V$ colours clearly indicate that the star-formation
has truncated. The `proto-arm' components ($\cal A$ and $\cal C$) have bluer and
younger colours, the offset in $U-B$ and less so in $B-V$ indicating these
regions are dustier. Even with just optical $UBVI$ imaging it is possible to
constrain the ages of morphologically resolved stellar populations while allowing
for dust and metallicity uncertainties. Resolved deep NIR imaging has enormous
potential for improving upon this type of analysis and will allow this `morphophotometric'
work to be highly constrained at $z>1$.
}
\end{figure}

It is highly desirable to be able to extend such work in the
NIR both because of the extended colour baseline to enable accurate
determination of stellar populations and to track the rest-frame
galaxy properties to high-redshifts. The ideal instrument for such
investigations would be a multi-band camera on a 8\,m telescope
equipped with Adaptive Optical (AO) correction to deliver high-resolution.
A wide field is not needed due to the high surface-density of
$z>1$ galaxies, in fact with many detectors it would be better to
splice in colour ({\em i.e.} using dichroics to cover many bands
simultaneously) than to cover a large area as AO correction is
limited by the size of the isoplanatic patch. 

Using such a camera in the
UV--NIR many `super-HDFs' could be observed and this
`morphophotometric' work extended to $z\sim 4$. A related
technology being developed at the AAO \cite{RUGATE}
is the concept of Rugate filters (Figure~5)
to allow broad-band observations in the NIR while still filtering out
night sky OH light. The Rugate transmission functions are multiply
peaked and can be tuned to allow light through only in OH-free regions.
This is achieved by stacking many hundreds of layers of different
refractive indices and thickness, the mathematical 
problem of ideal design, subject
to constraints of cost, has been solved
at AAO and enable $S/N$ gain factors of $\sim 2$
to be realised. This noise-reduction technique will allow us
to go deeper at a fraction of the cost of building a new telescope
with a larger aperture.

\begin{figure}
\psfig{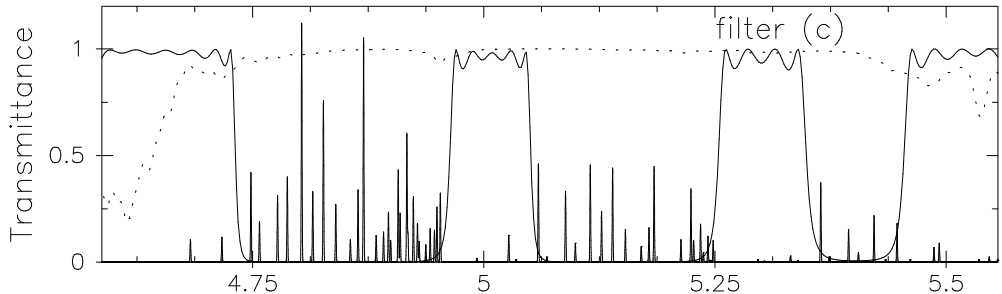}
\caption{\protect\small Sample Rugate filter design from the work of
Offer and Bland-Hawthorn. This shows the transmission function through
the J-band (solid-line) compared with the atmospheric transmission (dotted)
line and night-sky OH emission.
}
\end{figure}

\section{Spectroscopic Surveys}

While the signatures of evolution are clear enough in the deep imaging surveys the
determination of the {\em type} of evolution, {\it i.e.} whether it is in luminosity of
space density or something more complicated, requires redshift information in order to
construct luminosity functions. Redshifts are 
most reliably obtained by spectroscopy ---
determination by non-spectroscopic methods is subject to great uncertainty. For example
see Figure~6 which shows two independent estimates of photometric redshifts for faint
galaxies in the HDF in comparison with subsequent spectroscopy.

Spectroscopic surveys in the optical have determined luminosity function evolution out to
$z=1$ (Figure~6) --- to go beyond and probe the $z=1$--4 regime we need to do
multi-object NIR spectroscopy as the key spectroscopic features are redshifted in to the
$J$-band. 

Background suppression is even more critical for dispersed light and two
principal techniques have been proposed to go deep in the NIR:

\begin{enumerate}

\item {\bf OH Suppression}. Light can be dispersed in a spectrograph at high resolution 
($R=\lambda/\Delta\lambda \simeq 4000$), filtered through a mask
which is ruled to block light at the position of OH lines in the
spectrum, and undispersed (generally by reflecting back through the
same optical path) to form an image of the slit. This image can be 
then be detected (giving a very narrow background suppressed image)
or sent on through a conventional low-resolution ($R\sim 500$) IR
spectrograph to give an OH-suppressed spectrum. Such instruments 
are being built by several groups (e.g. Scaramella, these proceedings).
One disadvantage is that the extra optics can reduce the throughput by
a factor of two reducing the $S/N$ gain to only 2-3. Some sort of multi-object
arrangement can be had by use of fibres.

\item {\bf OH Avoidance}. The idea here is simply to build a high-resolution
$R=4000$ spectrograph and ignore the $\sim 10$\% of the pixels contaminated
by OH lines. The problem is the necessity of a large detector area to
get good wavelength coverage for redshift determination. For example three $1024^2$
NIR detectors would be required to cover the $J+H$ bands at this
resolution. Some designs introduce
cross-dispersing echelle type elements to reformat for one detector, although
again this introduces extra optics. It is extremely important to control
scattered light --- if it is not reduced below the 1\% level the supposed
gain is lost. This is by no means trivial to design for. Proposed designs
include CIRPASS (Parry \etal\ in Cambridge) which will have
30 Integral-Field fibre units positionable over a 10 arcmin field of view and the 
400 fibre multiobject AUSTRALIS design by Taylor (AAO) and Colless (MSSSO) 
to operate on the ESO VLT.

\end{enumerate} 

With such instruments coming online in the next 3--4 years we can
forsee measurement of the galaxy luminosity function out
to redshifts $z=1$--4. It is also possible to use current
NIR spectrographs --- for example \cite{KGB96} has been using 
CGS4 on the UKIRT to measure \Halpha\ in galaxies at $z\sim 1$.
By observing sources with {\em known} redshift, it is possible
to choose them so \Halpha\ avoids the OH lines, and the limited wavelength
range at high-resolution is not as important. 

Many of the proposed instruments feature Integral-Field Units, 
which opens the possibility of resolved spectroscopy of $z=1$--4
galaxies and the prospect of more detailed mapping of stellar populations,
than is possible with just broad-band colours, as well as measuring quantities such
as Tully-Fisher and $D_{\hbox{$n$}}-\sigma$.

\begin{figure}

\begin{tabular}{cc}
\psfig{file=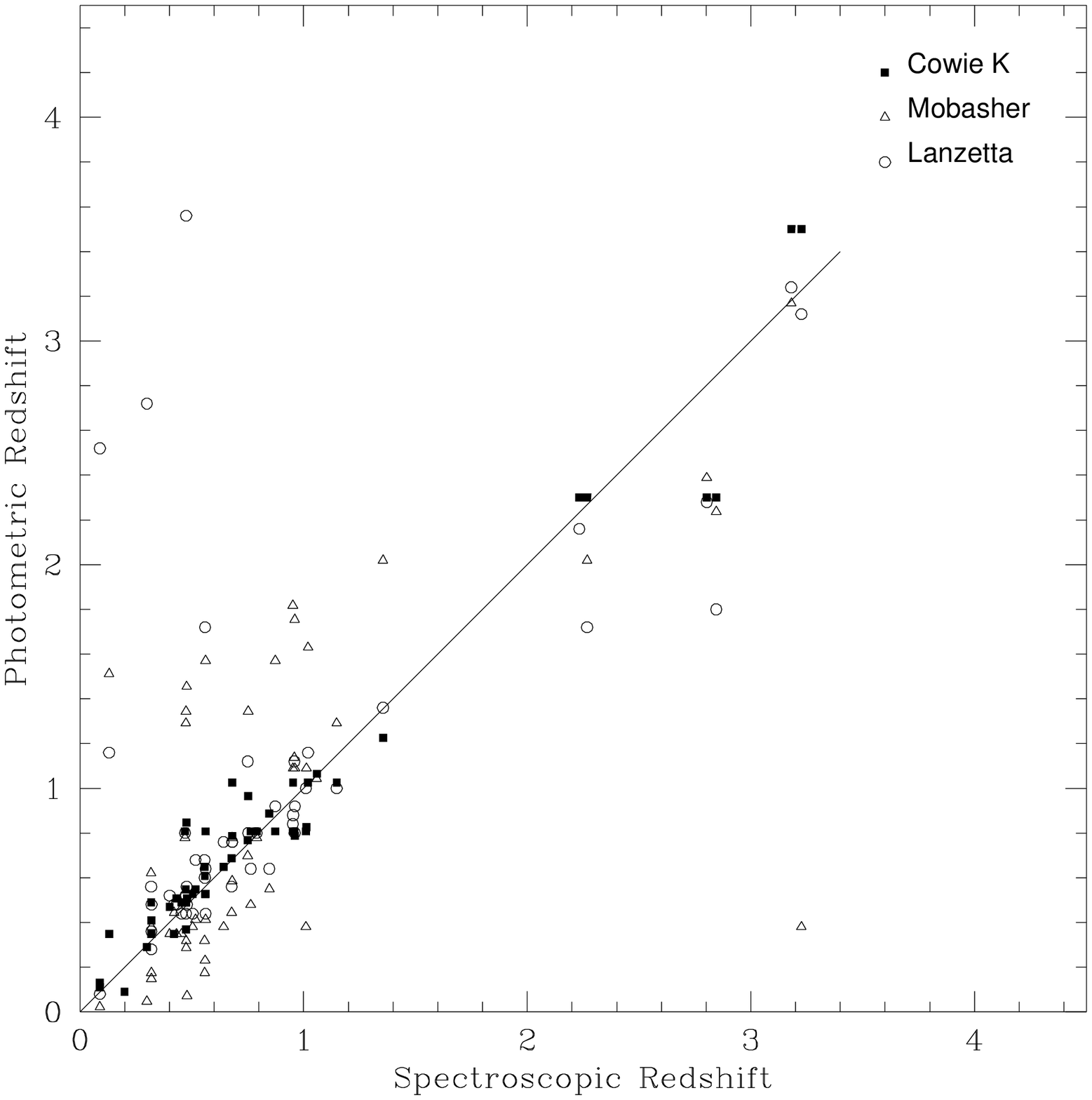,width=8cm} & 
\psfig{file=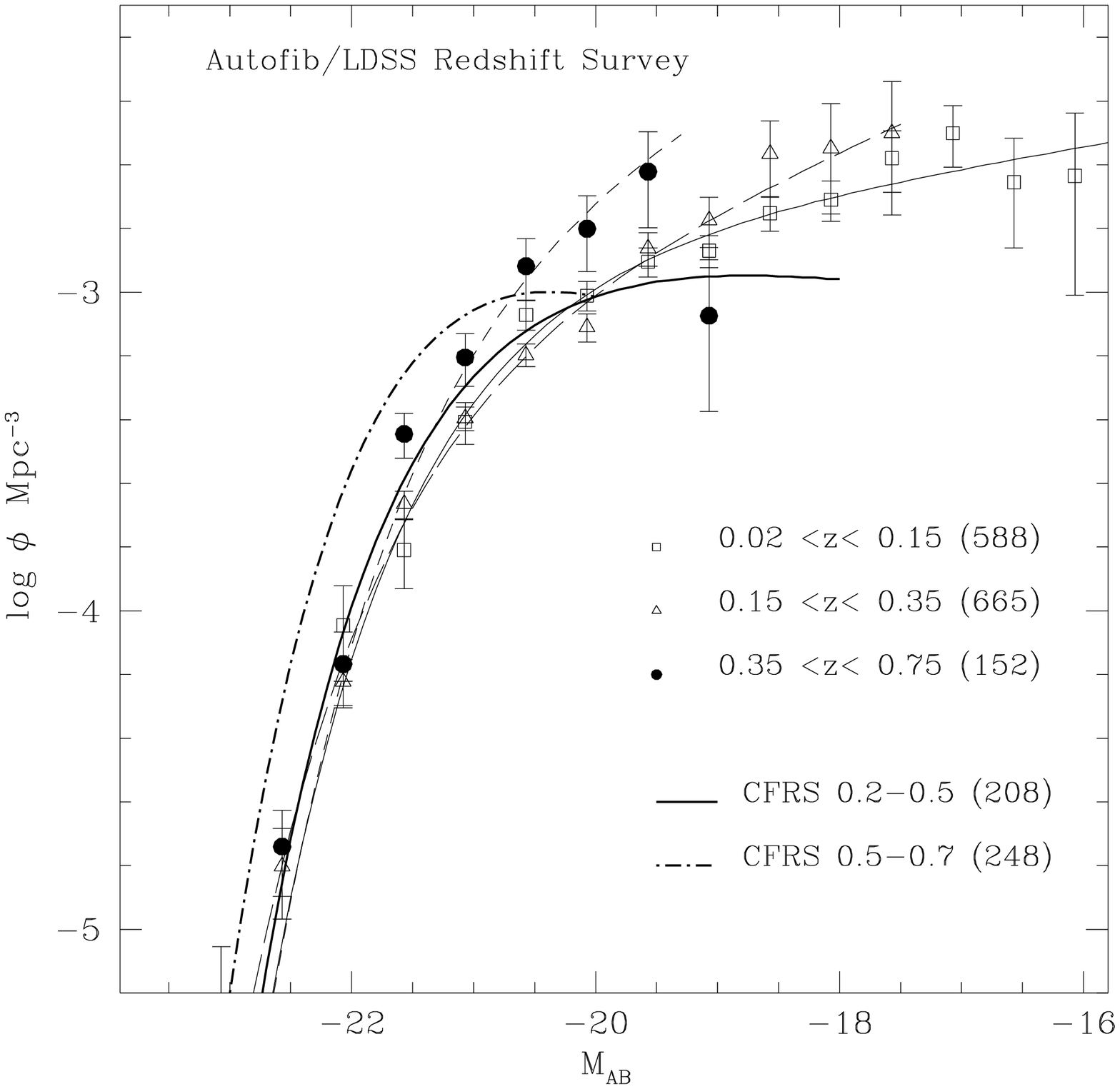,width=8cm}\\
\end{tabular}

\caption{\protect\small Left: Comparison of photometric and spectroscopic 
redshifts in the
HDF \protect\cite{COWIE96},
\protect\cite{LANZETTA}, \protect\cite{MOBASHER}. Photometric
redshifts exhibit a broad scatter, especially for blue galaxies where the flat
SED can fit a wide range of redshifts.\hfill\break
Right: The luminosity function evolution to $z=1$ determined
from
the AFRS \protect\cite{AFRS} and CFRS \protect\cite{CFRSVI} redshift surveys. The
primary evolutionary
effect is in the space density of galaxies at $M_B\sim -20$.}
\end{figure}

\section{Narrow-Band Imaging}

Rotation of the problem space by 90$^\circ$ leads us from 3D
spectroscopy to narrow-band imaging. There have been many
surveys looking for emission line galaxies at high-redshift,
using fixed narrow-band filters, in both
the optical (e.g. \cite{DEPROP}) and recently in 
the NIR (e.g. \cite{THO}) which have
found extreme objects but are not sensitive enough to reveal
normal star-forming populations at $z>1$.

At AAO we have a developed a new approach to narrow-band imaging 
using the Taurus Tunable Filter \cite{TTF}. The goal is
to provide a narrower filter which is better matched to the
typical emission-line widths of galaxies ($R\sim 1000$) than are
fixed glass interference filters ($R\sim 100$). This gives
the maximum $S/N$ contrast against sky in the line.
Moreover the wavelength
and resolution of the TTF is adjustable and can be scanned over the range.
This has been accomplished using a Fabry-Perot type technology --- the
key developments are: 

\begin{enumerate} 

\item Being able to reduce the plate gap down to
2--15\,\micron\ compared to 100--200\,\micron\ in a conventional Fabry-Perot.
The latter gives resolutions of $R\sim 10,000$ which is too {\em high}
to match galaxy lines. 

\item Large scan range --- TTF works from R=$100$--1000 so {\em all} uses
of conventional narrow band filters are superseded.

\item Improved Anti-Reflection coatings which enable the TTF to be used
over a {\em wide} bandpass. The current TTF is optimised for
use from 6000-10,000$\rm\,\AA$. A blue TTF which will be optimised for
work from
3500--$6000\rm\,\AA$ is on order and will be available at the AAT
in 1998.

\end{enumerate}

The TTF is being used for numerous projects, one that is interest
here is the TTF survey for $z>1$ emission line galaxies in the
Hubble Deep Field by myself, working with Abraham and Bland-Hawthorn. 
Estimates of star-formation rates (SFR) at low redshift and high redshift
indicate \cite{MADAU} that the SFR of the universe may peak
at $z=1.5$ (Figure~7) and this agrees with semi-empirical models \protect\cite{FALL}. 
Our survey is designed to test these this idea.
The key parameters of the survey are:

\begin{figure}[t]
\hskip 3.5cm\psfig{file=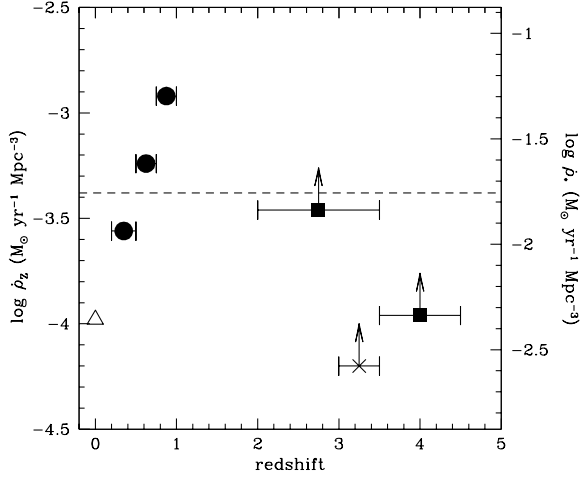,width=8cm}
\caption{\protect\small Star-formation density limits from the compilation
of \protect\cite{MADAU}. The points are: triangles --- local \Halpha\
survey \protect\cite{GALLEGO}, filled dots --- the CFRS
survey\protect\cite{CFRSXIII}, 
diagonal cross --- lower limits from Lyman limit galaxies observed by
\protect\cite{STEIDEL96}, filled squares --- from HDF \protect\cite{MADAU}.
The dashed line shows the average rate required to produce the current
abundance of metals for $\Omega=1$ and $H_0=50$ km\,s$^{-1}$\,Mpc$^{-1}$.}
Note: This diagram should be treated with caution, there
are many uncertainties to do with differing methods of measuring SFR,
large luminosity function corrections, etc. In this author's
opinion it should be regarded as a `first draft' of the history of SFR
in the universe. Independent measurements 
from diverse techniques based upon larger surveys will add considerably
to this picture in the next few years.
\end{figure}

\begin{enumerate}

\item We scan for [OII] emission over $0.9<z<1.5$ in 3 OH-line free  regions of the
$I$-band: I1: $z=0.894$--0.908 (0.2 hr$/$slice), I5: $z=$1.173--1.191 (1 hr$/$slice) and
I8: $z=$1.426--1.448, (1.5 hr$/$slice) for a total of 26 hours exposure. The exposure
times were chosen so as to go down to approximately similar line luminosities in the
different bands.  We covered an area of sky of 24 arcmin$^2$, which is a superset of the
entire HDF.

\item The survey was done using the TAURUS system on the William Herschel 4\,m
Telescope in La Palma, the HDF being a bit too far north for the AAT!
We had 4 clear nights in March 1997, with an average of $0.7$ arcsec seeing.

\item We aim to establish number counts, of limits thereof, of emission
line sources over $0.9<z<1.5$.
Redshifts we derive from our emission line sources will be
fed back into our morphological studies \cite{HDF}.

\end{enumerate}

Some sample data is shown in Figure~8. We are still working on a careful
analysis of our March dataset but we believe that tunable filter narrow-band
searches will be a productive new technique for finding emission line 
sources at high-redshift. By careful choice of wavelength the method of
OH avoidance is easy to practise, we are working on the design of a NIR
TTF device which will be the first to tackle the problem of tunable
imaging over a {\em large} field (e.g. 0.5 degree on a 8\,m at $f/2$
prime focus). Curvature of the plate surfaces can be used to combat
phase and non-telecentricity effects and achieve close to the ideal
monochromator design. Such a device
will allow us to extend such searches out into the $K$-band,
and pick up important lines such as \Halpha\ at $z=2$ \cite{NEVERHAPPEN}.

\begin{figure}
\psfig{file=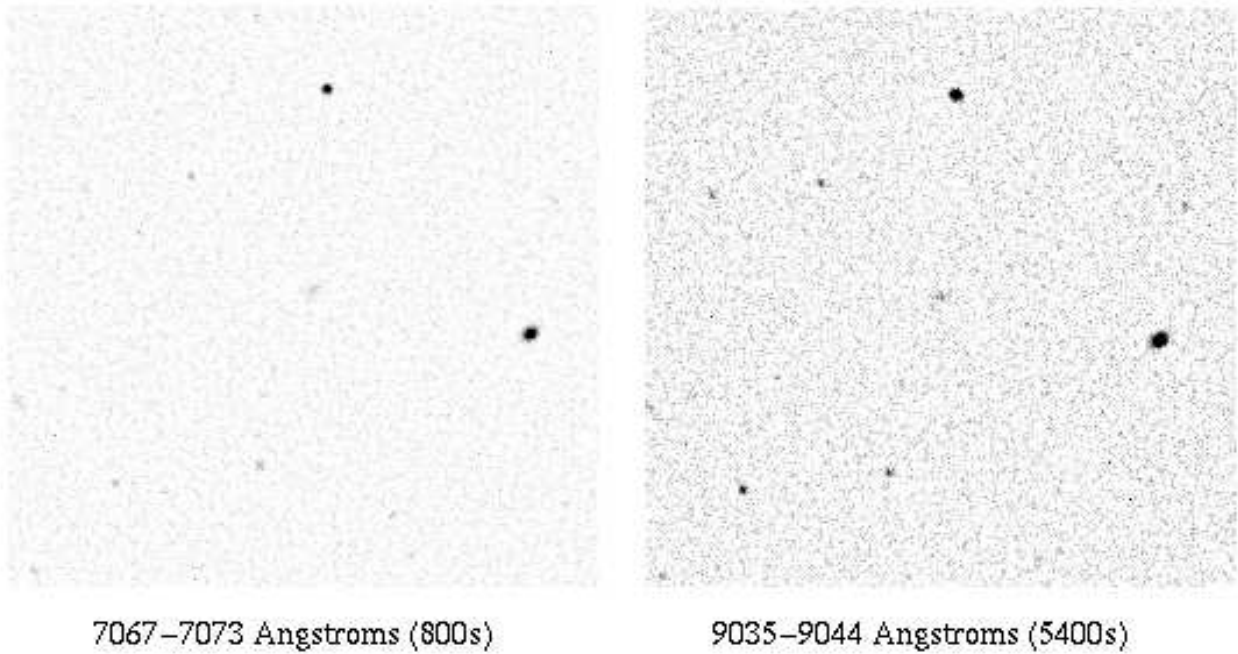,width=\hsize}
\caption{\protect\small
Deep narrow band images of the Hubble Deep Field (this section is 1.2 arcmin
on the side) taken with the Taurus Tunable Filter on the WHT in March,
1997, as part of a search for [OII] emission over the z-range 0.9-1.4.
The longer exposure of the 9040A slice allows for the lower CCD QE and
cosmological dimming of [OII]. These images show about 1/200 of our
data volume.
We are currently engaged in searching for emission line sources
in these sequences of images which should detect/constrain
the SFR over $z=0.9-1.4$.}
\end{figure}

\section{Summary}

It is clear that much progress has been made in the last few years in
understanding galaxy evolution at $z<1$, primarily driven by optical
studies. We now desire to track this evolution into the $z=1$--4 regime,
where preliminary evidence is mounting that galaxies are living in `interesting
times' and to do this we must move our observational methods in to the near
infrared. JHK imaging and spectroscopy accesses the rest-frame optical
and allows a direct comparison between the local and the distant
universe. The night sky brightness is a particular problem at these wavelengths,
but a variety of innovative instrument concepts are being proposed to 
reduce this background, and should allow $z=1$--4 galaxies to be accessed
with the next generation of 8\,m telescopes.

\acknowledgements{I am very grateful to many people, too numerous to mention,
for many discussions over the last few years on the subject of galaxy
evolution and instrumentation. I would particularly like to thank Bob
Abraham for providing figures from his latest work for this review, 
and Joss Bland-Hawthorn,
Keith Taylor and Ian Parry for introducing me to 
their innovative instrumental concepts.
I would also like to warmly thank the organisers of the Moriond conference
for inviting and funding me to give this review and the ski instructors
at Les Arcs for many painful bruises!}

\vfill
\end{document}